\documentclass[pra,aps,showpacs,groupedaddress,superscriptaddress,twocolumn,toc=flat]{revtex4-1}
\usepackage[colorlinks=true,linkcolor=blue,urlcolor=blue,citecolor=blue]{hyperref}

\usepackage[utf8x]{inputenc}
\usepackage{color}
\usepackage{bbm} 

\usepackage{amsfonts,amsmath,amssymb,stmaryrd}

\usepackage{graphicx}
\usepackage{subfigure}  
\usepackage{bbm} 
\usepackage{hyperref}
\usepackage{epsfig}
\usepackage{mathrsfs}
\usepackage{verbatim}
\usepackage{centernot}
\usepackage{ulem}
\usepackage{xspace}
\usepackage[english, status=draft]{fixme}       
\fxusetheme{color}                              %

\newcommand{\ket}[1]{|#1\rangle}

\usepackage{array}

\usepackage{cancel,ifthen}
\newcommand{\cmnt}[2][NoInPuT]{\ifthenelse{\equal{#1}{NoInPuT}}{}{{\color{red}\sout{#1}}} {\color{blue} #2}}

\usepackage{bm}	
\renewcommand{\vec}[1]{\bm{#1}}

\begin{document}
\normalem	

\title{Dominant fifth-order correlations in doped quantum anti-ferromagnets}

\author{A. Bohrdt}
\affiliation{Department of Physics and Institute for Advanced Study, Technical University of Munich, 85748 Garching, Germany}
\affiliation{Munich Center for Quantum Science and Technology (MCQST), Schellingstr. 4, D-80799 M\"unchen, Germany}

\author{Y. Wang}
\affiliation{Department of Physics, Harvard University, Cambridge, Massachusetts 02138, USA}

\author{J. Koepsell}
\affiliation{Max-Planck-Institut f\"ur Quantenoptik, 85748 Garching, Germany}
\affiliation{Munich Center for Quantum Science and Technology (MCQST), Schellingstr. 4, D-80799 M\"unchen, Germany}

\author{M. K\'anasz-Nagy}
\affiliation{Max-Planck-Institut f\"ur Quantenoptik, 85748 Garching, Germany}

\author{E. Demler}
\affiliation{Department of Physics, Harvard University, Cambridge, Massachusetts 02138, USA}

\author{F. Grusdt}
\affiliation{Department of Physics and Arnold Sommerfeld Center for Theoretical Physics (ASC), Ludwig-Maximilians-Universit\"at M\"unchen, Theresienstr. 37, M\"unchen D-80333, Germany}
\affiliation{Munich Center for Quantum Science and Technology (MCQST), Schellingstr. 4, D-80799 M\"unchen, Germany}


\date{\today}

\begin{abstract}
Traditionally one and two-point correlation functions are used to characterize many-body systems. In strongly correlated quantum materials, such as the doped 2D Fermi-Hubbard system, these may no longer be sufficient because higher-order correlations are crucial to understanding the character of the many-body system and can be numerically dominant. Experimentally, such higher-order correlations have recently become accessible in ultracold atom systems. Here we reveal strong non-Gaussian correlations in doped quantum anti-ferromagnets and show that higher order correlations dominate over lower-order terms. We study a single mobile hole in the $t-J$ model using DMRG, and reveal genuine fifth-order correlations which are directly related to the mobility of the dopant. We contrast our results to predictions using models based on doped quantum spin liquids which feature significantly reduced higher-order correlations. Our predictions can be tested at the lowest currently accessible temperatures in quantum simulators of the 2D Fermi-Hubbard model. Finally, we propose to experimentally study the same fifth-order spin-charge correlations as a function of doping. This will help to reveal the microscopic nature of charge carriers in the most debated regime of the Hubbard model, relevant for understanding high-$T_c$ superconductivity.
\end{abstract}

\maketitle

\emph{Introduction.--}
High-temperature superconductors are prime examples of strongly correlated quantum matter. In these quasi-2D systems superconductivity arises when mobile dopants are introduced into a parent anti-ferromagnetic (AFM) compound \cite{Lee2006,Keimer2015}, but the detailed mechanism remains elusive. It is widely believed that the interplay of spin and charge degrees of freedom plays a central role for understanding the underlying physics at low doping and can be described theoretically by the Fermi-Hubbard or $t-J$ model \cite{Zhang1988}. The central goal of this letter is to demonstrate that key features of magnetic dressing of doped holes in the Fermi Hubbard model can be revealed by analyzing five point spin-charge correlation functions. Furthermore, such high order correlation functions are found to be larger than the lower order ones in the regime of low doping and low temperatures. 

\begin{figure}[b!]
\centering
\epsfig{file=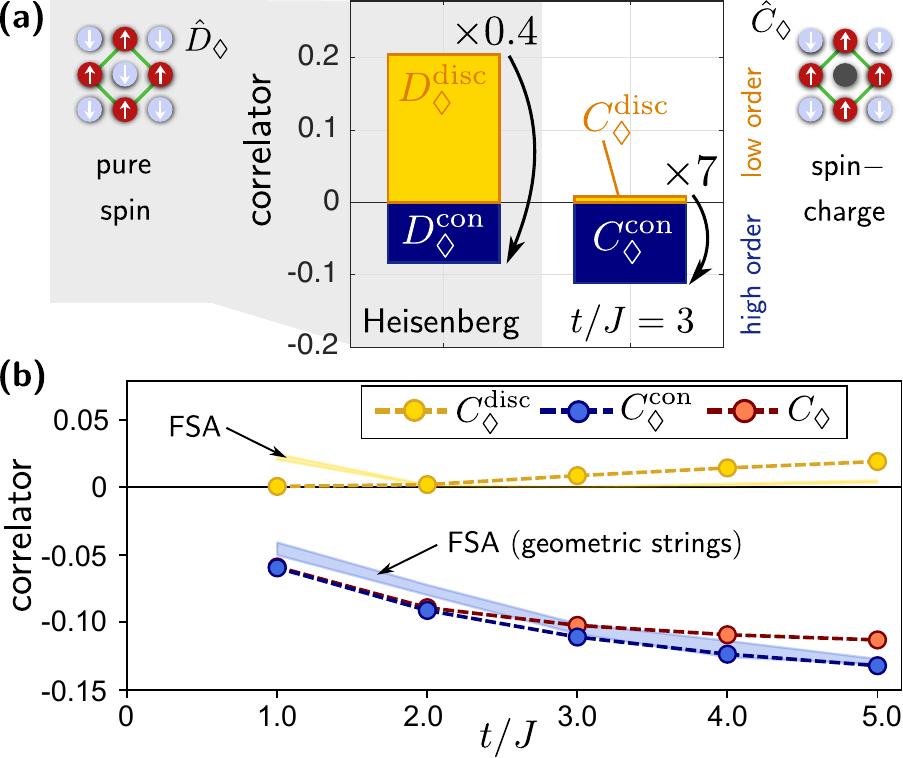, width=0.43\textwidth}
\caption{\textbf{Fifth-order spin-charge correlations} in a quantum AFM with a mobile dopant are studied in the 2D $t-J$ model and compared to the corresponding $4$th-order correlators in the undoped Heisenberg AFM (a); We use DMRG on a $6 \times 12$ cylinder and evaluate correlators at the center, in spin-balanced ensembles with $\langle \hat{S}^z \rangle = 0$. For one mobile dopant, genuine $5$-th order correlations $C_\lozenge^{\rm con}$ (Eq.~\eqref{eqCconnRedDef}, blue) are significantly larger (by a factor $\times 7$) than the lower-order disconnected terms $C_\lozenge^{\rm disc}=C_\lozenge-C_\lozenge^{\rm con}$ (yellow). In the undoped Heisenberg AFM the opposite is true: lower order correlators are dominant, while genuine higher-order correlations are smaller ($\times 0.4$). (b) Spin-charge correlators as a function of $t/J$. Our numerical results (data points, bare correlations $C_\lozenge$ in red) are explained by a frozen-spin approximation (FSA) ansatz (ribbons; width indicates statistical errors).} 
\label{fig1}
\end{figure}

Understanding the nature of charge carriers in strongly correlated electron systems, such as the doped Fermi Hubbard model, is a central problem of quantum many-body physics. While a single mobile hole inside the 2D quantum-Heisenberg AFM forms a magnetic (or spin-) polaron \cite{SchmittRink1988,Shraiman1988,Kane1989,Sachdev1989,Auerbach1991,Liu1991,Martinez1991,Brunner2000,Mishchenko2001,White2001,Koepsell2019,Grusdt2019PRB,Blomquist2019}, with spin and charge quantum numbers, it remains unknown whether spin and charge excitations (spinons and chargons) may become deconfined in the strange metal and pseudogap regimes, usually a characteristic of doped 1D spin chains \cite{Giamarchi2003,Kim1996,Sing2003,Vijayan2019}. Direct experimental or numerical evidence remains lacking so far. 

A common perspective on the puzzling properties of cuprates is the idea of several competing orders. Thus numerical studies of the Fermi Hubbard and $t-J$ models have often focused on the analysis of two point correlation functions with the goal of characterizing different types of broken symmetries. Furthermore, two point correlation functions can be naturally accessed in solid state systems using scattering experiments \cite{Birgeneau2006,Ament2011,Vig2015} and they play a central role in the development of effective mean-field theories. Recently, with the advent of quantum simulators based on ultracold atoms and ions, and especially quantum gas microscopes \cite{Bakr2009,Sherson2010,Parsons2015,Cheuk2015,Omran2015,Edge2015,Haller2015,Leibfried2003}, analysis of higher-order correlation functions, pioneered in Ref.~\cite{Schweigler2017}, has become a new experimental tool in the study of quantum many-body states \cite{Endres2011,Boll2016,Lukin2018a,Koepsell2019,Koepsell2020,Prufer2019,Zache2020}. They have rarely been studied so far, even though they contain a wealth of information about the underlying quantum states and are expected to become relevant when mean-field theories characterized by Gaussian correlations are no longer sufficient for capturing the physics.

\begin{figure}[t!]
\centering
\epsfig{file=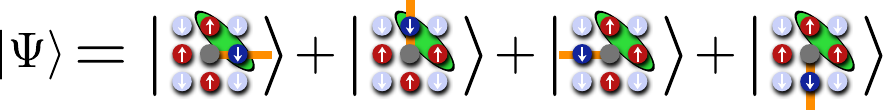, width=0.47\textwidth}
\caption{\textbf{The three-point spin-charge correlator} diagonally next to the hole, $\langle \hat{n}^{\rm h}_{\vec{j}} \hat{S}^z_{\vec{j} + \vec{e}_x} \hat{S}^z_{\vec{j}+\vec{e}_y}  \rangle$ (green bubble next to gray dot), vanishes for a mobile hole moving through a classical N\'eel state at the end of an infinitely long $S^z$-string. The fifth-order correlations $C_\lozenge \propto \langle \hat{n}^{\rm h}_{\vec{j}} \hat{S}^z_{\vec{j}+\vec{e}_x} \hat{S}^z_{\vec{j}+\vec{e}_y} \hat{S}^z_{\vec{j}-\vec{e}_x} \hat{S}^z_{\vec{j}-\vec{e}_y}  \rangle$ remain sizable and negative.} 
\label{fig2}
\end{figure}

Here, we identify lightly doped quantum AFMs as a system where genuine higher-order spin-charge correlations are present. Different from the situation in the undoped parent spin models, these higher-order correlations even dominate over lower-order terms when $t/J$ is sufficiently large, see Fig.~\ref{fig1} (a). Hence they provide a promising new diagnostic to unravel the nature of charge carriers and distinguish different theoretical models \cite{Huber2019}. Indeed, we show that even non-perturbative effective theories, such as doped resonating valence bond (RVB) states, cannot explain the higher-order correlations we find numerically at low doping. Instead, our results can be interpreted as signatures of geometric strings \cite{Grusdt2018PRX,Grusdt2019PRB} connecting spinons and chargons \cite{Beran1996,Laughlin1997,Baskaran2007,Sachdev2019}, reminiscent of an underlying $\mathbb{Z}_2$ Gauss law. Physically, strong five-point correlations indicate that a moving hole leaves behind a string of flipped spins, as shown in Fig.~\ref{fig2}.

In this paper, after introducing the higher-order spin-charge correlators, we perform DMRG ground state simulations of one mobile dopant on an extended cylinder. We then compare the latter to effective theories based on doped RVB states and the string picture. Finally, we use exact diagonalization to determine the temperature required for an experimental verification. In a follow-up paper we include the effect of a pinning potential \cite{Wang2020prep}.

\emph{Higher-order correlators.--}
We consider the following fifth-order ring spin-charge correlator,
\begin{equation}
C_\lozenge(\vec{r}) = \frac{2^4}{\langle \hat{n}^{\rm h}_{\mathbf{r}}\rangle } \langle \hat{n}^{\rm h}_{\vec{r}} \hat{S}^z_{\vec{r}+\vec{e}_x} \hat{S}^z_{\vec{r}+\vec{e}_y} \hat{S}^z_{\vec{r}-\vec{e}_x} \hat{S}^z_{\vec{r}-\vec{e}_y} \rangle,
\end{equation}
where $\hat{n}^{\rm h}_{\vec{r}}$ is the hole (dopant) density at site $\vec{r}$ and $\hat{S}^z_{\vec{j}}$ denotes the spin operator in $z$-direction at site $\vec{j}$. 
To witness the presence of genuine higher-order correlations, we calculate the connected correlator in the co-moving frame with the hole (defined in the supplements). In a spin-balanced ensemble with $\langle \hat{S}^z \rangle = 0$ (see discussion in Ref.~\cite{Wang2020prep}), expectation values with an odd number of $\hat{S}^z$ operators vanish and we obtain
\begin{equation}
C^{\rm con}_\lozenge = C_\lozenge - 2^4 \sum_{(\vec{i}, \vec{j}) \notin (\vec{k}, \vec{l})} 
  \frac{\langle \hat{n}^{\rm h}_{\vec{r}} \hat{S}^z_{\vec{r} + \vec{i}} \hat{S}^z_{\vec{r} + \vec{j}} \rangle}{\langle \hat{n}^{\rm h}_{\mathbf{r}}\rangle} 
 \frac{\langle \hat{n}^{\rm h}_{\vec{r}} \hat{S}^z_{\vec{r} + \vec{k}} \hat{S}^z_{\vec{r} + \vec{l}} \rangle}{\langle \hat{n}^{\rm h}_{\mathbf{r}}\rangle}.
 \label{eqCconnRedDef}
\end{equation}

In weakly correlated quantum systems, the values of higher-order correlation functions are dominated by more fundamental lower-order correlators. I.e. connected $n$-th order correlation functions $C^{\rm con}_n$ decrease with decreasing $n$, $|C^{\rm con}_1| > |C^{\rm con}_2| > ...$; For classical (product) states all connected correlations vanish $C^{\rm con}_n=0$, while in Gaussian systems only $C^{\rm con}_1, C^{\rm con}_2 \neq 0$ are non-zero \cite{Bruus2004}.

\emph{Magnetic polarons.--}
Intriguingly, for a magnetic polaron, formed when a single mobile hole is doped into an AFM, we find that the disconnected contributions from the lower-order correlators, $C^{\rm disc}_\lozenge = C_\lozenge - C_\lozenge^{\rm con}$, are significantly \emph{smaller} in magnitude than the higher-order correlators:  $| C^{\rm disc}_\lozenge | > |C_\lozenge(\vec{r})|, |C^{\rm con}_\lozenge(\vec{r})|$. In Fig.~\ref{fig1} we show DMRG results \cite{Note1}. for the ground state of a single hole in the $t-J$ \cite{Auerbach1998} model, as a function of $t/J$. The mobility of the dopant plays an important role for observing sizable higher-order spin-charge correlations. As $t/J$ is increased from $t/J=1$ to $t/J=5$, the absolute value of $C^{\rm con}_\lozenge$ approximately doubles. Throughout, the product of the lower-order two-point correlation functions is almost an order of magnitude smaller. 

We can define related $4$th-order correlators in the absence of doping as $D_\lozenge(\vec{r}) = 2^4 \langle \hat{S}^z_{\vec{r}+\vec{e}_x} \hat{S}^z_{\vec{r}+\vec{e}_y} \hat{S}^z_{\vec{r}-\vec{e}_x} \hat{S}^z_{\vec{r}-\vec{e}_y} \rangle$; a corresponding expression as in Eq.~\eqref{eqCconnDef} is obtained for the connected part $D^{\rm con}_\lozenge(\vec{r})$. In the classical Ising AFM, $D_\lozenge = 1$ and $D^{\rm con}_\lozenge = 0$. For the 2D Heisenberg model we performed DMRG simulations on a $6 \times 12$ cylinder and obtain $D_\lozenge = 0.12$ and $D^{\rm con}_\lozenge = - 0.083$ in the ground state, as indicated in Fig.~\ref{fig1} (a). The connected $4$th-order correlator only becomes negative because we subtract the significantly larger and positive two-point correlators, $D^{\rm disc} = D_\lozenge - D^{\rm con}_\lozenge = 0.20$. As expected for a weakly correlated quantum system, and different from the ground state with a mobile hole in the $t-J$ model, the lower-order correlators dominate in the 2D Heisenberg model: They are more than twice as large as the genuine fourth order correlations.

Orders of magnitude and the signs of $D^{(\rm con)}_\lozenge$ can be understood from a simple model of spontaneous symmetry breaking. Consider an ensemble of classical N\'eel states with AFM order parameters pointing in random directions. Because $D_\lozenge$ is always measured in the $z$-basis, we average over the entire ensemble and obtain
\begin{equation}
D_\lozenge |_{\rm cl} = 0.2, \qquad D^{\rm con}_\lozenge |_{\rm cl} = -0.133;
\label{eqC0NeelCl}
\end{equation}
These correlations are purely classical. Quantum fluctuations are expected to further reduce these values in the $SU(2)$ invariant Heisenberg model, as confirmed by our DMRG calculations.

\emph{Geometric and $S^z$- strings.--}
Negative $5$th-order correlations $C^{\rm con}_\lozenge(\vec{r}) < 0$ provide a signature of AFM correlations hidden by the motion of dopants. To understand the origin of such higher-order correlations, we first consider a toy model of a single hole in an Ising AFM pointing along the $S^z$- direction. Neglecting string configurations affecting more than one spin in the direct vicinity of the mobile dopant, we notice that $C_\lozenge$ switches sign if the hole is attached to a string $\Sigma$ of over-turned spins ($S^z$-string) of length $\ell>0$. Hence $C_\lozenge$ can be expressed by the probability $p_{\ell>0}$ for the string to have non-zero length, namely $C_\lozenge \approx p_{\ell=0} - p_{\ell>0}$ or $ C_\lozenge \approx 1-2 p_{\ell>0}$. 

Assuming that the system is in an equal superposition of all string configurations, we can estimate various correlation functions. Because the hole is equally likely to occupy either sublattice, $\langle  \hat{n}^{\rm h}_{\vec{r}}  \hat{S}^z_{\vec{r}\pm\vec{e}_{x,y}} \rangle = 0$. Three-point correlations $\langle  \hat{n}^{\rm h}_{\vec{r}}  \hat{S}^z_{\vec{r}\pm\vec{e}_{x,y}}  \hat{S}^z_{\vec{r}\pm\vec{e}_{x,y}} \rangle = 0$ vanish, as can be seen by averaging over the four possible orientations of the first string segment, counting from the hole, and neglecting string configurations which affect more than one spin in the immediate vicinity of the hole, see Fig.~\ref{fig2}. Hence, Eq.~\eqref{eqCconnDef} implies $C^{\rm con}_\lozenge(\vec{r}) = C_\lozenge(\vec{r}) \approx -1$ for sufficiently many non-zero strings $p_{\ell>0} \approx 1$. 

In this setting, relevant to the 2D $t-J_z$ model \cite{Chernyshev1999,Grusdt2018PRX}, $C_\lozenge$ takes the role of a $\mathbb{Z}_2$ Gauss law: the mobile dopants represent $\mathbb{Z}_2$ charges and the $\mathbb{Z}_2$ electric field lines correspond to $S^z$-strings of overturned spins. Similarly, in the $SU(2)$ invariant $t-J$ or Fermi-Hubbard models the higher-order correlator $C_\lozenge$ serves as an indicator for geometric strings \cite{Grusdt2018PRX,Grusdt2019PRB} of displaced spins. 

The $t/J$ dependence observed in Fig.~\ref{fig1} can be explained within the geometric string theory by a frozen-spin approximation (FSA) ansatz \cite{Grusdt2018PRX}. As in Refs.~\cite{Chiu2019Science,Bohrdt2019NatPhys} we start from snapshots of the Heisenberg ground state in the Fock basis along $S^z$ and create a hole by randomly removing one spin. This dopant is subsequently moved through the system in random directions, re-arranging the positions of the surrounding spins while keeping their quantum state frozen; the string length distribution is calculated from a linear string model with string tension $dE/d\ell = 2 J (C_2 - C_1)$ \cite{Grusdt2018PRX}, where $C_{1(2)}$ are nearest (next-nearest) neighbor spin correlations in the undoped AFM. This way, new sets of snapshots are generated for every value of $t/J$, from which the higher-order correlators can then be obtained \cite{Chiu2019Science}.

\begin{figure}[t!]
\centering
\epsfig{file=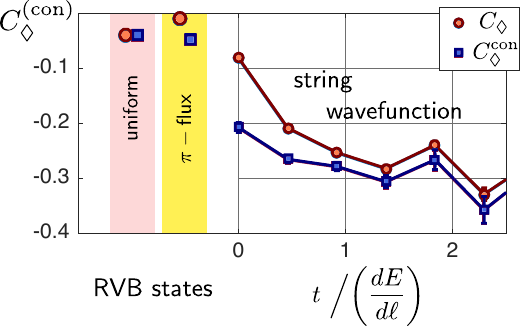, width=0.4\textwidth} ~~
\caption{\textbf{Comparison of RVB and geometric string trial states} in a $14 \times 14$ system with $S^z_{\rm tot}=1/2$. For the 'plain vanilla' uniform and $\pi$-flux RVB states doped with a single hole (left), $C_\lozenge^{(\rm con)}$ (from Eq.~\eqref{eqCconnDef}) is small. The string wavefunction (right), with a weak $SU(2)$ breaking staggered magnetization along $S^z$, exhibits larger values of the spin-charge correlator and shows a strong dependence on the ratio of $t$ and the string tension $dE/d\ell =1.09 J$ \cite{Grusdt2019PRB} which determines the average length of geometric strings in the trial state. Note that the doped RVB states have no $t/J$ dependence.} 
\label{fig4}
\end{figure}

\emph{Doped spin liquids.--}
A class of microscopic variational wavefunctions that has been used to model doped quantum spin liquids is based on Anderson's RVB paradigm \cite{Anderson1987,Baskaran1987}. Being able to resolve properties of the many-body wavefunction on microscopic scales, ultracold atom experiments provide an opportunity to put the RVB theory to a rigorous experimental test in a clean system.

Here we calculate the higher-order spin charge correlations $C_\lozenge^{(\rm con)}$ for two paradigmatic doped RVB trial states. The uniform RVB state starts from an unpolarized Fermi sea $\ket{\rm FS}$ of free spin-up and spin-down spinons $\hat{f}_{\vec{k},\sigma}$. To describe a free hole excitation moving through the system, one spinon with momentum $\vec{k}$ and spin $\sigma$ is removed. A meaningful trial state for the $t-J$ model, without double occupancies and independent of $t/J$, is obtained by applying the Gutzwiller projection: $ \ket{\Psi_{\rm uRVB}} = \mathcal{N} \hat{\mathcal{P}}_{\rm GW} \hat{f}_{\vec{k},\sigma} \ket{\rm FS}$ normalized by $\mathcal{N}$. We use standard Metropolis Monte Carlo sampling \cite{Gros1989} to evaluate $C_\lozenge^{(\rm con)}$ in the trial state $\ket{\Psi_{\rm uRVB}}$, and show our result in Fig.~\ref{fig4}. We find $C_\lozenge^{(\rm con)} = -0.040(4)$ with significantly smaller magnitude than found for large values of $t/J$ by DMRG, cf. Fig.~\ref{fig1}.

We find a similar result for the doped $\pi$-flux RVB state \cite{Giamarchi1993}, for which decent agreement with experimental data has recently been reported in ultracold atoms at finite doping \cite{Bohrdt2019NatPhys,Chiu2019Science}. The $\pi$-flux state with a single hole has the same form as the $\rm uRVB$ state above, except that the Fermi sea $\ket{\rm FS}$ is replaced by a Dirac semi-metal of spinons obtained when introducing $\pi$ magnetic flux per plaquette in the effective spinon Hamiltonian \cite{Wen1996}. In this case $C_\lozenge^{\rm con} = -0.049(3)$ slightly increases and $C_\lozenge = -0.008(3)$ decreases in magnitude. Both are significantly weaker than numerically expected from DMRG when $t > J$.

In a recently proposed extension of the RVB ansatz, geometric strings are included in the trial wavefunction \cite{Grusdt2019PRB,Bohrdt2020ARPES}. Now we demonstrate that the presence of such geometric strings increases the expected higher-order correlators. We start from the optimized RVB wavefunction for half filling (no doping) \cite{Lee1988,Piazza2015}, which includes a weak spontaneously formed staggered magnetization along $S^z$-direction. A spinon is removed in the usual way and after the Gutzwiller projection a geometric string is added to the hole, thus re-arranging the spins surrounding the dopant; see Refs.~\cite{Grusdt2019PRB,Bohrdt2020ARPES} for details. 

In Fig.~\ref{fig4} (right panel) we show how $C_\lozenge^{(\rm con)}$ evaluated for this string wavefunction depends on the ratio of hole tunneling $t$ and the linear string tension $dE/d\ell$ underlying the model. When $t/J=0$ the length of geometric strings is zero and the observed increase of the higher-order correlator is due to the staggered magnetization along $S^z$ included in the trial wavefunction. For increasing tunneling $t$ the string length grows and another significant increase of $C_\lozenge^{(\rm con)}$ is observed. This supports our picture that the mobility of dopants leads to long geometric strings, which in turn underly strong higher-order spin-charge correlations. 

\begin{figure}[t!]
\centering
\epsfig{file=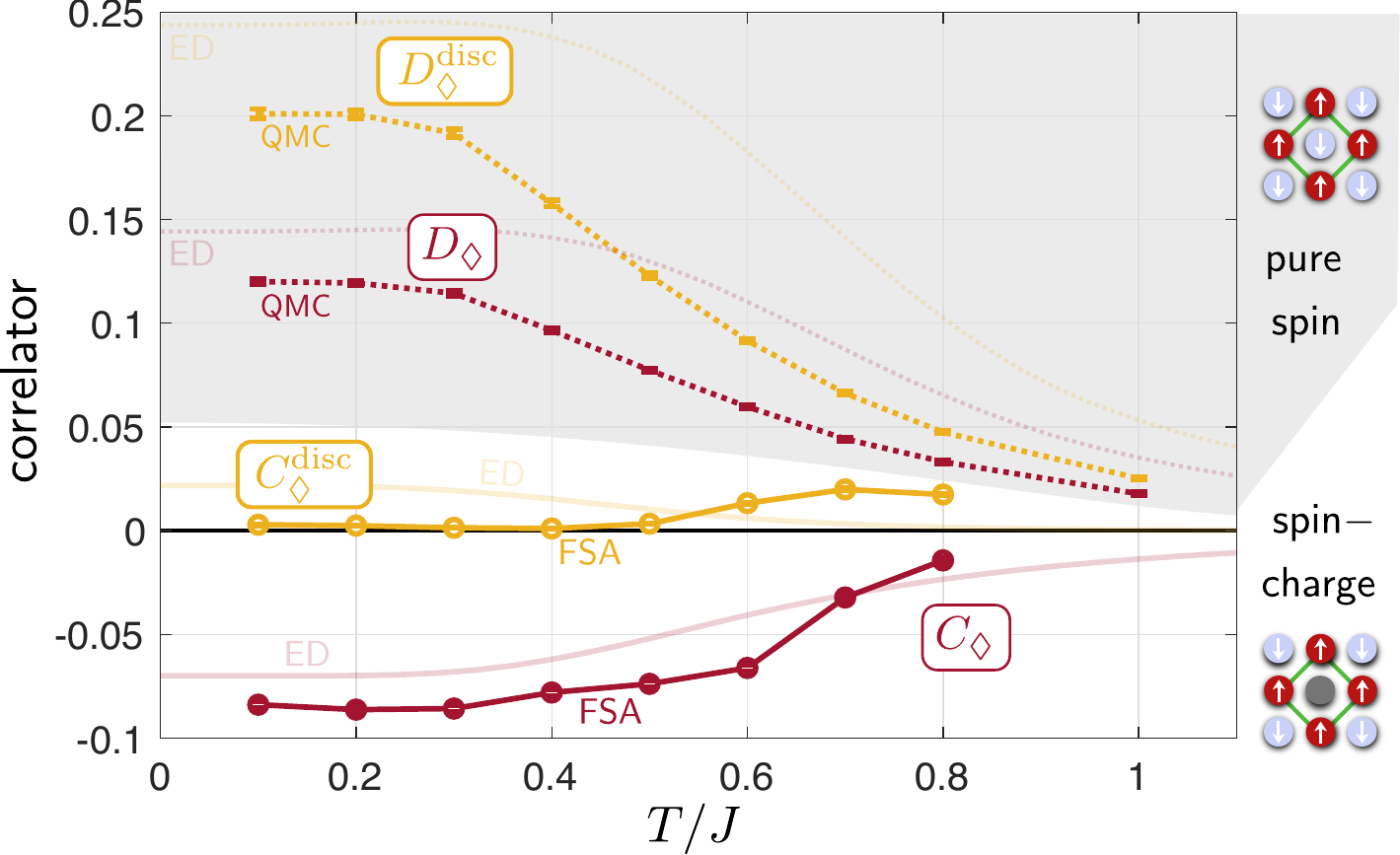, width=0.499\textwidth}
\caption{\textbf{Temperature dependence} of the higher-order correlations $C_\lozenge$ (red) and the disconnected $C_\lozenge^{\rm disc}$ (yellow) parts (using Eq.~\eqref{eqCconnRedDef}). We compare the corresponding correlators $D_\lozenge$ in the undoped Heisenberg model (top) to predictions for a mobile dopant (with $t/J = 2$), using the FSA based on Heisenberg quantum Monte Carlo (QMC) snapshots and ED simulations in a periodic $4 \times 4$ system. 
} 
\label{fig5}
\end{figure}

\emph{Experimental considerations.--}
We turn to a discussion of the limitations and requirements to observe higher-order correlations in the doped Hubbard model.

Fig.~\ref{fig5} demonstrates how thermal fluctuations suppress higher-order correlations. We show the lower-order disconnected terms $C^{\rm disc}_\lozenge$ and compare them to the higher-order correlators $C_\lozenge$, in two cases: For the undoped Heisenberg model we use quantum Monte Carlo simulations \cite{Note2}. For a single mobile dopant our predictions are based on the frozen-spin approximation (FSA / geometric strings) \cite{Grusdt2018PRX,Bohrdt2019NatPhys,Chiu2019Science} described above; The correlators are evaluated from $10^4$ snapshots for each $T/J$. We also compare to exact diagonalization (ED) calculations in a $4 \times 4$ system and find good agreement.

In the geometric string theory, $C^{\rm disc}_\lozenge$ is approximately zero up to temperatures $T\lesssim 0.5J$, while the higher-order correlator $C_\lozenge$ is of the order of $-0.1$. Without a hole, the disconnected part $D^{\rm disc}_\lozenge$ is significantly larger than $D_\lozenge$ for these small temperatures. For $T \gtrsim  0.6J$ the correlations decay quickly. The relevant temperature range has already been accessed experimentally \cite{Mazurenko2017}, and we expect that more quantum gas microscopes operating in this regime will follow in the near future \cite{Chiu2018PRL,Boll2016,Cheuk2015,Brown2017,Edge2015}. Similar to quantum gas microscopy, the higher-order correlations in Fig.~\ref{fig5} are extracted from snapshots. We expect that the main experimental challenge will be to collect sufficient amounts of data to obtain acceptable errorbars. Current experiments offering simultaneous spin- and charge resolution \cite{Koepsell2020} are very close to the temperature regime required for observing the higher-order correlations proposed here.

\emph{Summary and Outlook.--}
We propose to study fifth-order spin-charge correlations to explore the microscopic nature of charge carriers in the doped Hubbard model from a new perspective. Such correlators have recently become accessible in state-of-the-art quantum gas microscopes. The observables we consider are direct generalizations of the three-point spin-charge correlators $\langle \hat{S}^z_{j-1} \hat{n}^{\rm h}_j \hat{S}^z_{j+1} \rangle$ underlying hidden AFM correlations in the 1D doped Hubbard model \cite{Kruis2004a,Hilker2017}. We analyze similar correlations in 2D, which can only be understood by theories with non-Gaussian correlations.

Our numerical studies for a single doped hole reveal the importance of the hole mobility for establishing such higher-order correlations and making them become the dominant spin-charge correlations in the system. In a subsequent work, we demonstrate this explicitly by considering the effect of a localized pinning potential for the hole \cite{Wang2020prep}. Here we also established that doped quantum spin liquids have reduced higher-order correlations, whereas fluctuating geometric strings can explain the observed enhancement when $t/J$ is increased. 

An interesting question concerns the behavior of the higher-order correlations when doping is increased and numerical studies of the Fermi-Hubbard model become more challenging \cite{LeBlanc2015}. In this regime we propose to measure the higher-order correlators by state-of-the-art ultracold atom experiments. Such studies can shed new light on the nature of charge carriers in the pseudogap and strange-metal \cite{Brown2019a} regimes or the pairing mechanism between dopants. They also provide a new experimental route to distinguish theoretical trial states, e.g. in the RVB class. While a recent machine-learning analysis \cite{Bohrdt2019NatPhys} suggests that up to $\simeq 15 \%$ doping a model based on geometric strings may be favorable compared to doped $\pi$-flux RVB states, further refined experiments as proposed here will be required to establish where and how the nature of charge carriers changes upon doping. 

Our results can be applied to extend studies of the formation dynamics of magnetic polarons \cite{Bohrdt2019Dyn,Hubig2020,Ji2020}, or to investigate correlation effects in Bose polaron problems in an optical lattice \cite{Fukuhara2013}. Other possible extensions include the study of $SU(2)$ invariant generalizations of the higher-order spin-charge correlators introduced here.

\emph{Acknowledgements.--}
The authors would like to thank T. Hilker, T. Zache, M. Pr\"ufer, M. Knap, D. Sels, P. Preiss, S. Jochim and I. Bloch for useful discussions. We acknowledge funding by the Deutsche Forschungsgemeinschaft (DFG, German Research Foundation) under Germany's Excellence Strategy -- EXC-2111 -- 390814868. Y.W. acknowledges the Postdoctoral Fellowship of the Harvard-MPQ Center for Quantum Optics and AFOSR-MURI Award No. FA95501610323. This research used resources of the National Energy Research Scientific Computing Center (NERSC), a U.S. Department of Energy Office of Science User Facility operated under Contract No. DE-AC02- 05CH11231. J.K. acknowledges funding from Hector Fellow Academy and supported by the Max Planck Harvard Research Center for Quantum Optics (MPHQ). ED and YW acknowledge support from Harvard-MIT CUA, Harvard-MPQ Center, ARO grant number W911NF-20-1-0163, and the National Science Foundation through grant No. OAC-1934714.

~\\
\newpage

\begin{figure}[t!]
\centering
\epsfig{file=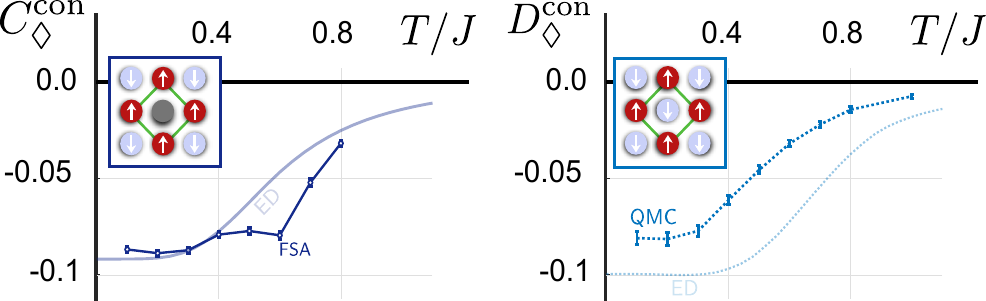, width=0.48\textwidth}
\caption{For the data shown in Fig.~\ref{fig5} of the main text, we plot the connected correlators $C_\lozenge^{\rm con}$ (left) and $D_\lozenge^{\rm con}$ (right).
} 
\label{figSM1}
\end{figure}

\begin{widetext}
\section*{Supplements}
We define the connected correlator in the co-moving frame with the hole as,
\begin{multline}
C^{\rm con}_\lozenge(\vec{r}) = C_\lozenge(\vec{r}) - 2^4  \biggl[ \sum_{\vec{l} \notin (\vec{i},\vec{j},\vec{k})} 
\frac{ \langle \hat{n}^{\rm h}_{\mathbf{r}} \hat{S}^z_{\vec{r} + \vec{i}} \hat{S}^z_{\vec{r} + \vec{j}} \hat{S}^z_{\vec{r} + \vec{k}}\rangle_c}{\langle \hat{n}^{\rm h}_{\mathbf{r}}\rangle} 
 \frac{\langle \hat{n}^{\rm h}_{\vec{r}} \hat{S}^z_{\vec{r} + \vec{l}} \rangle}{{\langle \hat{n}^{\rm h}_{\mathbf{r}}\rangle}}
+ \sum_{(\vec{i}, \vec{j}) \notin (\vec{k}, \vec{l})} 
  \frac{\langle \hat{n}^{\rm h}_{\vec{r}} \hat{S}^z_{\vec{r} + \vec{i}} \hat{S}^z_{\vec{r} + \vec{j}} \rangle_c}{\langle \hat{n}^{\rm h}_{\mathbf{r}}\rangle} 
 \frac{\langle \hat{n}^{\rm h}_{\vec{r}} \hat{S}^z_{\vec{r} + \vec{k}} \hat{S}^z_{\vec{r} + \vec{l}} \rangle_c}{\langle \hat{n}^{\rm h}_{\mathbf{r}}\rangle} \\
+ \sum_{\vec{i} \neq \vec{j} \notin (\vec{k}, \vec{l})}  
 \frac{ \langle \hat{n}^{\rm h}_{\vec{r}} \hat{S}^z_{\vec{r} + \vec{i}} \rangle }{\langle \hat{n}^{\rm h}_{\mathbf{r}}\rangle}
 \frac{ \langle \hat{n}^{\rm h}_{\vec{r}} \hat{S}^z_{\vec{r} + \vec{j}} \rangle }{\langle \hat{n}^{\rm h}_{\mathbf{r}}\rangle}
 \frac{ \langle  \hat{n}^{\rm h}_{\vec{r}} \hat{S}^z_{\vec{r} + \vec{k}} \hat{S}^z_{\vec{r} + \vec{l}} \rangle_c }{\langle \hat{n}^{\rm h}_{\mathbf{r}}\rangle} 
+ \frac{ \langle  \hat{n}^{\rm h}_{\vec{r}}  \hat{S}^z_{\vec{r}+\vec{e}_x} \rangle }{\langle \hat{n}^{\rm h}_{\mathbf{r}}\rangle} 
  \frac{ \langle  \hat{n}^{\rm h}_{\vec{r}}  \hat{S}^z_{\vec{r}+\vec{e}_y} \rangle  }{\langle \hat{n}^{\rm h}_{\mathbf{r}}\rangle}
  \frac{ \langle  \hat{n}^{\rm h}_{\vec{r}}  \hat{S}^z_{\vec{r}-\vec{e}_x} \rangle  }{\langle \hat{n}^{\rm h}_{\mathbf{r}}\rangle}
  \frac{ \langle  \hat{n}^{\rm h}_{\vec{r}}  \hat{S}^z_{\vec{r}-\vec{e}_y} \rangle }{\langle \hat{n}^{\rm h}_{\mathbf{r}}\rangle}
\biggr],
\label{eqCconnDef}
\end{multline}
where the sums are over disjoint sets of lower-order connected correlators (defined equivalently) involving sites $\vec{i}, \vec{j}, \vec{k}, \vec{l} = \pm \vec{e}_{x,y}$ measured relative to the hole. In a spin-balanced ensemble (see discussion in Ref.~\cite{Wang2020prep}), the expectation values $\langle \hat{n}^{\rm h}_{\vec{r}} \hat{S}^z_{\vec{r} + \vec{l}} \rangle = 0$ vanish and we obtain Eq.~\eqref{eqCconnRedDef}.

In Fig.~\ref{figSM1} we show the connected correlators $C_\lozenge^{\rm con}$ corresponding to the data from Fig.~\ref{fig5}.
\end{widetext}

\begin{thebibliography}{10}

\bibitem{Lee2006}
Patrick~A. Lee, Naoto Nagaosa, and Xiao-Gang Wen.
\newblock Doping a mott insulator: Physics of high-temperature
  superconductivity.
\newblock {\em Rev. Mod. Phys.}, 78:17--85, Jan 2006.

\bibitem{Keimer2015}
B.~Keimer, S.~A. Kivelson, M.~R. Norman, S.~Uchida, and J.~Zaanen.
\newblock From quantum matter to high-temperature superconductivity in copper
  oxides.
\newblock {\em Nature}, 518:179--, February 2015.

\bibitem{Zhang1988}
F.~C. Zhang and T.~M. Rice.
\newblock Effective hamiltonian for the superconducting cu oxides.
\newblock {\em Phys. Rev. B}, 37:3759--3761, Mar 1988.

\bibitem{SchmittRink1988}
S.~Schmitt-Rink, C.~M. Varma, and A.~E. Ruckenstein.
\newblock Spectral function of holes in a quantum antiferromagnet.
\newblock {\em Phys. Rev. Lett.}, 60:2793--2796, Jun 1988.

\bibitem{Shraiman1988}
Boris~I. Shraiman and Eric~D. Siggia.
\newblock Mobile vacancies in a quantum heisenberg antiferromagnet.
\newblock {\em Phys. Rev. Lett.}, 61:467--470, Jul 1988.

\bibitem{Kane1989}
C.~L. Kane, P.~A. Lee, and N.~Read.
\newblock Motion of a single hole in a quantum antiferromagnet.
\newblock {\em Phys. Rev. B}, 39:6880--6897, Apr 1989.

\bibitem{Sachdev1989}
Subir Sachdev.
\newblock Hole motion in a quantum n\'eel state.
\newblock {\em Phys. Rev. B}, 39:12232--12247, Jun 1989.

\bibitem{Auerbach1991}
Assa Auerbach and Brond~E. Larson.
\newblock Small-polaron theory of doped antiferromagnets.
\newblock {\em Phys. Rev. Lett.}, 66:2262--2265, Apr 1991.

\bibitem{Liu1991}
Zhiping Liu and Efstratios Manousakis.
\newblock Spectral function of a hole in the t-j model.
\newblock {\em Phys. Rev. B}, 44:2414--2417, Aug 1991.

\bibitem{Martinez1991}
Gerardo Martinez and Peter Horsch.
\newblock Spin polarons in the t-j model.
\newblock {\em Phys. Rev. B}, 44:317--331, Jul 1991.

\bibitem{Brunner2000}
Michael Brunner, Fakher~F. Assaad, and Alejandro Muramatsu.
\newblock Single-hole dynamics in the $t\ensuremath{-}j$ model on a square
  lattice.
\newblock {\em Phys. Rev. B}, 62:15480--15492, Dec 2000.

\bibitem{Mishchenko2001}
A.~S. Mishchenko, N.~V. Prokof'ev, and B.~V. Svistunov.
\newblock Single-hole spectral function and spin-charge separation in the $t-j$
  model.
\newblock {\em Phys. Rev. B}, 64:033101, Jun 2001.

\bibitem{White2001}
Steven~R. White and Ian Affleck.
\newblock Density matrix renormalization group analysis of the nagaoka polaron
  in the two-dimensional $t\ensuremath{-}j$ model.
\newblock {\em Phys. Rev. B}, 64:024411, Jun 2001.

\bibitem{Koepsell2019}
Joannis Koepsell, Jayadev Vijayan, Pimonpan Sompet, Fabian Grusdt, Timon~A.
  Hilker, Eugene Demler, Guillaume Salomon, Immanuel Bloch, and Christian
  Gross.
\newblock Imaging magnetic polarons in the doped fermi-hubbard model.
\newblock {\em Nature}, 572(7769):358--362, 2019.

\bibitem{Grusdt2019PRB}
Fabian Grusdt, Annabelle Bohrdt, and Eugene Demler.
\newblock Microscopic spinon-chargon theory of magnetic polarons in the
  $t\text{\ensuremath{-}}j$ model.
\newblock {\em Phys. Rev. B}, 99:224422, Jun 2019.

\bibitem{Blomquist2019}
Emil Blomquist and Johan Carlström.
\newblock Ab initio description of magnetic polarons in a mott insulator.
\newblock {\em arXiv:1912.08825}.

\bibitem{Giamarchi2003}
Thierry Giamarchi.
\newblock {\em Quantum Physics in One Dimension}.
\newblock Oxford University Press, 2003.

\bibitem{Kim1996}
C.~Kim, A.~Y. Matsuura, Z.-X. Shen, N.~Motoyama, H.~Eisaki, S.~Uchida,
  T.~Tohyama, and S.~Maekawa.
\newblock Observation of spin-charge separation in one-dimensional
  srcu${\mathrm{o}}_{2}$.
\newblock {\em Phys. Rev. Lett.}, 77:4054--4057, Nov 1996.

\bibitem{Sing2003}
M.~Sing, U.~Schwingenschl\"ogl, R.~Claessen, P.~Blaha, J.~M.~P. Carmelo, L.~M.
  Martelo, P.~D. Sacramento, M.~Dressel, and C.~S. Jacobsen.
\newblock Electronic structure of the quasi-one-dimensional organic conductor
  ttf-tcnq.
\newblock {\em Phys. Rev. B}, 68:125111, Sep 2003.

\bibitem{Vijayan2019}
Jayadev Vijayan, Pimonpan Sompet, Guillaume Salomon, Joannis Koepsell, Sarah
  Hirthe, Annabelle Bohrdt, Fabian Grusdt, Immanuel Bloch, and Christian Gross.
\newblock Time-resolved observation of spin-charge deconfinement in fermionic Hubbard chains.
\newblock {\em Science}, 367, 186--189 (2020).

\bibitem{Birgeneau2006}
R.~J. Birgeneau, C.~Stock, J.~M. Tranquada, and K.~Yamada.
\newblock Magnetic neutron scattering in hole doped cuprate superconductors.
\newblock {\em Journal of the Physical Society of Japan}, 75, 111003 (2006) 

\bibitem{Ament2011}
Luuk J.~P. Ament, Michel van Veenendaal, Thomas~P. Devereaux, John~P. Hill, and
  Jeroen van~den Brink.
\newblock Resonant inelastic x-ray scattering studies of elementary
  excitations.
\newblock {\em Rev. Mod. Phys.}, 83:705--767, Jun 2011.

\bibitem{Vig2015}
Sean Vig, Anshul Kogar, Matteo Mitrano, Ali~A. Husain, Vivek Mishra, Melinda~S.
  Rak, Luc Venema, Peter~D. Johnson, Genda~D. Gu, Eduardo Fradkin, Michael~R.
  Norman, and Peter Abbamonte.
\newblock Measurement of the dynamic charge response of materials using
  low-energy, momentum-resolved electron energy-loss spectroscopy (M-EELS).
  \newblock {\em SciPost Phys.}, 3, 026 (2017).

\bibitem{Bakr2009}
Waseem~S. Bakr, Jonathon~I. Gillen, Amy Peng, Simon Foelling, and Markus
  Greiner.
\newblock A quantum gas microscope for detecting single atoms in a
  hubbard-regime optical lattice.
\newblock {\em Nature}, 462(7269):74--U80, November 2009.

\bibitem{Sherson2010}
Jacob~F. Sherson, Christof Weitenberg, Manuel Endres, Marc Cheneau, Immanuel
  Bloch, and Stefan Kuhr.
\newblock Single-atom-resolved fluorescence imaging of an atomic mott
  insulator.
\newblock {\em Nature}, 467(7311):68--U97, September 2010.

\bibitem{Parsons2015}
Maxwell~F. Parsons, Florian Huber, Anton Mazurenko, Christie~S. Chiu, Widagdo
  Setiawan, Katherine Wooley-Brown, Sebastian Blatt, and Markus Greiner.
\newblock Site-resolved imaging of fermionic $^{6}\mathrm{Li}$ in an optical
  lattice.
\newblock {\em Phys. Rev. Lett.}, 114:213002, May 2015.

\bibitem{Cheuk2015}
Lawrence~W. Cheuk, Matthew~A. Nichols, Melih Okan, Thomas Gersdorf, Vinay~V.
  Ramasesh, Waseem~S. Bakr, Thomas Lompe, and Martin~W. Zwierlein.
\newblock Quantum-gas microscope for fermionic atoms.
\newblock {\em Phys. Rev. Lett.}, 114:193001, May 2015.

\bibitem{Omran2015}
Ahmed Omran, Martin Boll, Timon~A. Hilker, Katharina Kleinlein, Guillaume
  Salomon, Immanuel Bloch, and Christian Gross.
\newblock Microscopic observation of pauli blocking in degenerate fermionic
  lattice gases.
\newblock {\em Phys. Rev. Lett.}, 115:263001, Dec 2015.

\bibitem{Edge2015}
G.~J.~A. Edge, R.~Anderson, D.~Jervis, D.~C. McKay, R.~Day, S.~Trotzky, and
  J.~H. Thywissen.
\newblock Imaging and addressing of individual fermionic atoms in an optical
  lattice.
\newblock {\em Phys. Rev. A}, 92:063406, Dec 2015.

\bibitem{Haller2015}
Elmar Haller, James Hudson, Andrew Kelly, Dylan~A. Cotta, Bruno Peaudecerf,
  Graham~D. Bruce, and Stefan Kuhr.
\newblock Single-atom imaging of fermions in a quantum-gas microscope.
\newblock {\em Nat Phys}, 11(9):738--742, September 2015.

\bibitem{Leibfried2003}
Dietrich Leibfried, Rainer Blatt, Christopher Monroe, and David Wineland.
\newblock Quantum dynamics of single trapped ions.
\newblock {\em Reviews of Modern Physics}, 75(1):281, 2003.

\bibitem{Schweigler2017}
T.~ Schweigler, V.~ Kasper, S.~ Erne, I.~ Mazets, B.~ Rauer, F.~ Cataldini, T.~ Langen, T.~ Gasenzer, J.~ Berges, and J\"org Schmiedmayer.
\newblock Experimental characterization of a quantum many-body system via higher-order correlations.
\newblock {\em Nature}, 545, 323 - 326 (2017).

\bibitem{Endres2011}
M.~Endres, M.~Cheneau, T.~Fukuhara, C.~Weitenberg, P.~Schauss, C.~Gross,
  L.~Mazza, M.~C. Banuls, L.~Pollet, I.~Bloch, and S.~Kuhr.
\newblock Observation of correlated particle-hole pairs and string order in
  low-dimensional mott insulators.
\newblock {\em Science}, 334(6053):200--203, 2011.

\bibitem{Boll2016}
Martin Boll, Timon~A. Hilker, Guillaume Salomon, Ahmed Omran, Jacopo Nespolo,
  Lode Pollet, Immanuel Bloch, and Christian Gross.
\newblock Spin- and density-resolved microscopy of antiferromagnetic
  correlations in fermi-hubbard chains.
\newblock {\em Science}, 353(6305):1257--1260, 2016.

\bibitem{Lukin2018a}
Alexander Lukin, Matthew Rispoli, Robert Schittko, M.~Eric Tai, Adam~M.
  Kaufman, Soonwon Choi, Vedika Khemani, Julian Léonard, and Markus Greiner.
\newblock Probing entanglement in a many-body-localized system.
\newblock {\em Science}, 364, 6437, 256-260 (2019).

\bibitem{Koepsell2020}
Joannis Koepsell, Sarah Hirthe, Dominik Bourgund, Pimonpan Sompet, Jayadev
  Vijayan, Guillaume Salomon, Christian Gross, and Immanuel Bloch.
\newblock Robust bilayer charge-pumping for spin- and density-resolved quantum
  gas microscopy.
\newblock {\em Phys. Rev. Lett.} 125, 010403 (2020).

\bibitem{Prufer2019}
Maximilian Prüfer, Torsten~V. Zache, Philipp Kunkel, Stefan Lannig, Alexis
  Bonnin, Helmut Strobel, Jürgen Berges, and Markus~K. Oberthaler.
\newblock Experimental extraction of the quantum effective action for a
  non-equilibrium many-body system.
\newblock {\em arXiv:1909.05120}.

\bibitem{Zache2020}
Torsten~V. Zache, Thomas Schweigler, Sebastian Erne, J\"org Schmiedmayer, and
  J\"urgen Berges.
\newblock Extracting the field theory description of a quantum many-body system
  from experimental data.
\newblock {\em Phys. Rev. X}, 10:011020, Jan 2020.

\bibitem{Huber2019}
Sebastian Huber, Fabian Grusdt, and Matthias Punk.
\newblock Signatures of correlated magnetic phases in the two-spin density
  matrix.
\newblock {\em Phys. Rev. A}, 99:023617, Feb 2019.

\bibitem{Grusdt2018PRX}
F.~Grusdt, M.~K\'anasz-Nagy, A.~Bohrdt, C.~S. Chiu, G.~Ji, M.~Greiner,
  D.~Greif, and E.~Demler.
\newblock Parton theory of magnetic polarons: Mesonic resonances and signatures
  in dynamics.
\newblock {\em Phys. Rev. X}, 8:011046, Mar 2018.

\bibitem{Beran1996}
P.~Beran, D.~Poilblanc, and R.B. Laughlin.
\newblock Evidence for composite nature of quasiparticles in the 2d t-j model.
\newblock {\em Nuclear Physics B}, 473(3):707--720, 1996.

\bibitem{Laughlin1997}
R.~B. Laughlin.
\newblock Evidence for quasiparticle decay in photoemission from underdoped
  cuprates.
\newblock {\em Phys. Rev. Lett.}, 79:1726--1729, Sep 1997.

\bibitem{Baskaran2007}
G.~Baskaran.
\newblock 3/2-fermi liquid: the secret of high-tc cuprates.
\newblock {\em arXiv:0709.0902}, 2007.

\bibitem{Sachdev2019}
Subir Sachdev, Harley~D. Scammell, Mathias~S. Scheurer, and Grigory
  Tarnopolsky.
\newblock Gauge theory for the cuprates near optimal doping.
\newblock {\em Phys. Rev. B}, 99:054516, Feb 2019.

\bibitem{Wang2020prep}
Y.~Wang, A.~Bohrdt, J.~Koepsell, F.~Grusdt, and E.~Demler.
\newblock {\em in preparation}.

\bibitem{Bruus2004}
Henrik Bruus and Karsten Flensberg.
\newblock {\em Many-Body Quantum Theory in Condensed Matter Physics}.
\newblock Oxford University Press, 2004.

\bibitem{Note1}
Our code uses the TeNPy package \cite {hauschildTenpy,Hauschild2018SciPost}. We
  averaged over the six legs of the cylinder to speed up convergence of the
  considered higher-order correlator.
  
\bibitem{hauschildTenpy}
J.~ Hauschild, R.~ Mong, F.~ Pollmann, M.~ Schulz, L.~ Schoonderwoert, J.~ Unfried, Y.~ Tzeng, and M.~ Zaletel.
\newblock {\em The code is available online at \url{https://github.com/tenpy/tenpy/}, the documentation can be found at \url{https://tenpy.readthedocs.io}}.
  
\bibitem{Hauschild2018SciPost}
 J.~ Hauschild, and F.~ Pollmann.
\newblock {\em Efficient numerical simulations with Tensor Networks: Tensor Network Python (TeNPy)}.
\newblock {\em SciPost Phys. Lect. Notes}, 5 (2018).

\bibitem{Auerbach1998}
Assa Auerbach.
\newblock {\em Interacting Electrons and Quantum Magnetism}.
\newblock Springer, Berlin, 1998.

\bibitem{Chernyshev1999}
A.~L. Chernyshev and P.~W. Leung.
\newblock Holes in the $t\ensuremath{-}{J}_{z}$ model: A diagrammatic study.
\newblock {\em Phys. Rev. B}, 60:1592--1606, Jul 1999.

\bibitem{Chiu2019Science}
Christie~S. Chiu, Geoffrey Ji, Annabelle Bohrdt, Muqing Xu, Michael Knap,
  Eugene Demler, Fabian Grusdt, Markus Greiner, and Daniel Greif.
\newblock String patterns in the doped hubbard model.
\newblock {\em Science}, 365(6450):251--256, 2019.

\bibitem{Bohrdt2019NatPhys}
Annabelle Bohrdt, Christie~S. Chiu, Geoffrey Ji, Muqing Xu, Daniel Greif,
  Markus Greiner, Eugene Demler, Fabian Grusdt, and Michael Knap.
\newblock Classifying snapshots of the doped hubbard model with machine
  learning.
\newblock {\em Nature Physics}, 15:921--924, 2019.

\bibitem{Anderson1987}
P.~W. Anderson.
\newblock The resonating valence bond state in la2cuo4 and superconductivity.
\newblock {\em Science}, 235(4793):1196--1198, 1987.

\bibitem{Baskaran1987}
G.~Baskaran, Z.~Zou, and P.W. Anderson.
\newblock The resonating valence bond state and high-tc superconductivity - a
  mean field theory.
\newblock {\em Solid State Communications}, 63(11):973--976, 1987.

\bibitem{Gros1989}
Claudius Gros.
\newblock Physics of projected wavefunctions.
\newblock {\em Annals of Physics}, 189(1):53--88, 1989.

\bibitem{Giamarchi1993}
T.~Giamarchi and C.~Lhuillier.
\newblock Dispersion relation of a single hole in the t-j model determined by a
  variational monte carlo method.
\newblock {\em Phys. Rev. B}, 47:2775--2779, Feb 1993.

\bibitem{Wen1996}
Xiao-Gang Wen and Patrick~A. Lee.
\newblock Theory of underdoped cuprates.
\newblock {\em Phys. Rev. Lett.}, 76:503--506, Jan 1996.

\bibitem{Bohrdt2020ARPES}
Annabelle Bohrdt, Eugene Demler, Frank Pollmann, Michael Knap, and Fabian
  Grusdt.
\newblock Parton theory of arpes spectra in anti-ferromagnetic mott insulators.
\newblock {\em arXiv:2001.05509}.

\bibitem{Lee1988}
T.~K. Lee and Shiping Feng.
\newblock Doping dependence of antiferromagnetism in
  ${\mathrm{la}}_{2}$cu${\mathrm{o}}_{4}$: A numerical study based on a
  resonating-valence-bond state.
\newblock {\em Phys. Rev. B}, 38:11809--11812, Dec 1988.

\bibitem{Piazza2015}
B.~Dalla~Piazza, M.~Mourigal, N.~B. Christensen, G.~J. Nilsen,
  P.~Tregenna-Piggott, T.~G. Perring, M.~Enderle, D.~F. McMorrow, D.~A. Ivanov,
  and H.~M. Ronnow.
\newblock Fractional excitations in the square-lattice quantum antiferromagnet.
\newblock {\em Nat Phys}, 11(1):62--68, January 2015.

\bibitem{Note2}
We used the same Monte-Carlo code as in Ref.~\cite {Chiu2019Science}.

\bibitem{Mazurenko2017}
Anton Mazurenko, Christie~S. Chiu, Geoffrey Ji, Maxwell~F. Parsons, Marton
  Kanasz-Nagy, Richard Schmidt, Fabian Grusdt, Eugene Demler, Daniel Greif, and
  Markus Greiner.
\newblock A cold-atom fermi-hubbard antiferromagnet.
\newblock {\em Nature}, 545(7655):462--466, May 2017.

\bibitem{Chiu2018PRL}
Christie~S. Chiu, Geoffrey Ji, Anton Mazurenko, Daniel Greif, and Markus
  Greiner.
\newblock Quantum state engineering of a hubbard system with ultracold
  fermions.
\newblock {\em Phys. Rev. Lett.}, 120:243201, Jun 2018.

\bibitem{Brown2017}
Peter~T. Brown, Debayan Mitra, Elmer Guardado-Sanchez, Peter Schauss,
  Stanimir~S. Kondov, Ehsan Khatami, Thereza Paiva, Nandini Trivedi, David~A.
  Huse, and Waseem~S. Bakr.
\newblock Spin-imbalance in a 2d fermi-hubbard system.
\newblock {\em Science}, 357(6358):1385--, September 2017.

\bibitem{Kruis2004a}
H.~V. Kruis, I.~P. McCulloch, Z.~Nussinov, and J.~Zaanen.
\newblock Geometry and the hidden order of luttinger liquids: The universality
  of squeezed space.
\newblock {\em Phys. Rev. B}, 70:075109, Aug 2004.

\bibitem{Hilker2017}
Timon~A. Hilker, Guillaume Salomon, Fabian Grusdt, Ahmed Omran, Martin Boll,
  Eugene Demler, Immanuel Bloch, and Christian Gross.
\newblock Revealing hidden antiferromagnetic correlations in doped hubbard
  chains via string correlators.
\newblock {\em Science}, 357(6350):484--487, 2017.

\bibitem{LeBlanc2015}
J.~P.~F. LeBlanc, Andrey~E. Antipov, Federico Becca, Ireneusz~W. Bulik, Garnet
  Kin-Lic Chan, Chia-Min Chung, Youjin Deng, Michel Ferrero, Thomas~M.
  Henderson, Carlos~A. Jim\'enez-Hoyos, E.~Kozik, Xuan-Wen Liu, Andrew~J.
  Millis, N.~V. Prokof'ev, Mingpu Qin, Gustavo~E. Scuseria, Hao Shi, B.~V.
  Svistunov, Luca~F. Tocchio, I.~S. Tupitsyn, Steven~R. White, Shiwei Zhang,
  Bo-Xiao Zheng, Zhenyue Zhu, and Emanuel Gull.
\newblock Solutions of the two-dimensional hubbard model: Benchmarks and
  results from a wide range of numerical algorithms.
\newblock {\em Phys. Rev. X}, 5:041041, Dec 2015.

\bibitem{Brown2019a}
Peter~T. Brown, Debayan Mitra, Elmer Guardado-Sanchez, Reza Nourafkan, Alexis
  Reymbaut, Charles-David Hebert, Simon Bergeron, A.-M.~S. Tremblay, Jure
  Kokalj, David~A. Huse, Peter Schau\ss, and Waseem~S. Bakr.
\newblock Bad metallic transport in a cold atom fermi-hubbard system.
\newblock {\em Science}, 363(6425):379--, January 2019.

\bibitem{Bohrdt2019Dyn}
Annabelle Bohrdt, Fabian Grusdt, and Michael Knap.
\newblock Dynamical formation of a magnetic polaron in a two-dimensional
  quantum antiferromagnet.
\newblock {\em arXiv:1907.08214}, 2019.

\bibitem{Hubig2020}
Claudius Hubig, Annabelle Bohrdt, Michael Knap, Fabian Grusdt, and Ignacio
  Cirac.
\newblock Evaluation of time-dependent correlators after a local quench in
  ipeps: hole motion in the t-j model.
\newblock {\em SciPost Phys.}, 8(2):021--, February 2020.

\bibitem{Ji2020}
Geoffrey Ji, Muqing Xu, Lev~Haldar Kendrick, Christie~S. Chiu, Justus~C.
  Brüggenjürgen, Daniel Greif, Annabelle Bohrdt, Fabian Grusdt, Eugene
  Demler, Martin Lebrat, and Markus Greiner.
\newblock Dynamical interplay between a single hole and a hubbard
  antiferromagnet.
\newblock {\em arXiv:2006.06672v1}.

\bibitem{Fukuhara2013}
Takeshi Fukuhara, Adrian Kantian, Manuel Endres, Marc Cheneau, Peter Schauss,
  Sebastian Hild, David Bellem, Ulrich Schollwoeck, Thierry Giamarchi,
  Christian Gross, Immanuel Bloch, and Stefan Kuhr.
\newblock Quantum dynamics of a mobile spin impurity.
\newblock {\em Nature Physics}, 9(4):235--241, April 2013.

\end{thebibliography}
\bibliographystyle{unsrt}

\end{document}